# New Insights in Boundary-only and Domain-type RBF Methods

## W. Chen[*] and M. Tanaka[**]

*Department of Mechanical System Engineering, Shinshu University, Wakasato 4-17-1, Nagano City, Nagano 380-8533, Japan*
(*E-mail:* [*] chenw@homer.shinshu-u.ac.jp *and* [**] dtanaka@gipwc.shinshu-u.ac.jp)

This paper has made some significant advances in the boundary-only and domain-type RBF techniques. The proposed boundary knot method (BKM) is different from the standard boundary element method in a number of important aspects. Namely, it is truly meshless, exponential convergence, integration-free (of course, no singular integration), boundary-only for general problems, and leads to symmetric matrix under certain conditions (able to be extended to general cases after further modified). The BKM also avoids the artificial boundary in the method of fundamental solution. An amazing finding is that the BKM can formulate linear modeling equations for nonlinear partial differential systems with linear boundary conditions. This merit makes it circumvent all perplexing issues in the iteration solution of nonlinear equations. On the other hand, by analogy with Green's second identity, we also presents a general solution RBF (GSR) methodology to construct efficient RBFs in the domain-type RBF collocation method and dual reciprocity method. The GSR approach first establishes an explicit relationship between the BEM and RBF itself on the ground of the potential theory. This paper also discusses some essential issues relating to the RBF computing, which include time-space RBFs, direct and indirect RBF schemes, finite RBF method, and the application of multipole and wavelet to the RBF solution of the PDEs.

## 1. Introduction

The radial basis function (RBF) is a powerful concept to numerical computations. For example, the introducing the RBF into the BEM [1] has eliminated its major weakness in handling inhomogeneous terms. Nevertheless, the constructing efficient RBFs is still an open research topic. This communication focuses on this problem.

First, we introduce the non-singular general solution as the RBF to derive a meshless, exponential convergence, integration-free, and boundary-only technique. The dual reciprocity method (DRM) and RBF are also employed jointly here to approximate particular solution of inhomogeneous terms as in dual reciprocity BEM (DRBEM) and the method of fundamental solution (MFS). This combined method circumvents either singular integration, slow convergence and mesh in the DRBEM [2] or artificial boundary outside physical domain in the MFS [2,3]. Additionally, the present method holds the symmetric matrix structure for self-adjoint operators subject to a kind of boundary condition. These merits lead to tremendous improvement in computational accuracy, efficiency and stability. The method is named as the boundary knot method (BKM) to differentiate it from the other numerical techniques. The term "boundary-only" is used here in the sense as in the DRBEM and MFS that only boundary knots are required, although internal knots can improve solution accuracy. It is also found that the BKM can produce linear modeling for nonlinear problems with nonlinear governing equations and linear boundary conditions [4]. Two numerical examples are provided to verify the efficiency and utility of this new technique. The completeness concern of the BKM is also discussed.

Second, the RBF has a very close tie with the indirect BEM. By analogy with Green's second identity, we present a new general solution RBFs (GSR) construction methodology for inhomogeneous problems in the domain-type collocation method and the DRM, which include superconvergent pre-wavelet RBFs. As an interesting extension of the normal RBF concept, the time-space RBFs (TSR) is introduced to deal with time variable equally as the space variables. The finite RBF method (FRM) is also defined, which employ the finite RBF support in a truncated manner like finite differences but without mesh. The FRM results in a symmetric sparse banded system matrix, which eliminates a major limitation of the global RBF methods. We also discuss the possible limitations of the indirect RBF schemes for problems with sharp corners and nonlinear terms.

## 2. Approximation of Particular Solution

The BKM can be viewed as a two-step numerical scheme, namely, approximation of particular solution and the evaluation of homogeneous solution. The former has been well developed in [2]. For completeness, we outline its basic methodology. Let us consider the differential equation

$$L\{u\} = f(x) + \rho\{u\} \quad (1)$$

defined in a region $V$ bounded by a surface $S$, where $L$ represents a differential operator with available non-singular general solution, $\rho\{u\}$ is the remaining differential operator. $f(x)$ is a known forcing function and $x$ means mulitdimension independent variable. Boundary conditions on $S$ may also be inhomogeneous:

$$u(x) = D(x), \quad x \subset S_u, \quad (2a)$$

$$\frac{\partial u(x)}{\partial n} = N(x), \quad x \subset S_T, \quad (2b)$$

where $n$ is the unit outward normal. The solution of Eq. (1) can be expressed as

$$u = v + u_p, \quad (3)$$

where $v$ and $u_p$ are the general and particular solutions, respectively. The latter satisfies equation

$$L_1\{u_p\} = f(x) + \rho\{u\}, \quad (4)$$

but does not necessarily satisfy boundary conditions. $v$ is the homogeneous solution

$$L\{v\} = 0, \quad (5)$$

$$v(x) = D(x) - u_p, \quad (6a)$$

$$\frac{\partial v(x)}{\partial n} = N(x) - \frac{\partial u_p(x)}{\partial n}, \quad (6b)$$

Eqs. (5) and (6a,b) will be later solved by the RBF using non-singular general solution. The DRM analogizes the particular solution by the use of a series of approximate particular solution at all specified nodes of boundary and domain. The inhomogeneous terms of Eq. (4) are approximated by

$$f(x) + \rho(u) \cong \sum_{j=1}^{N+L} \alpha_j \phi(r_j), \quad (7)$$

where $\alpha_j$ are the unknown coefficients. $N$ and $L$ are respectively the numbers of knots on boundary and domain. $r_j = \|x - x_j\|$ represents the Euclidean norm, $\phi$ is the RBF.

By forcing Eq. (7) to exactly satisfy Eq. (4) at all nodes, we can uniquely determine

$$\alpha = A_\phi^{-1}\left[f(x) + \rho\{A_\phi\}A_\phi^{-1}u\right], \quad (8)$$

where $A_\phi$ is nonsingular RBF interpolation matrix and $I$ the unite matrix.

Finally, we can get particular solutions at any point by summing localized approximate particular solutions

$$u_p = \sum_{j=1}^{N+L} \alpha_j \varphi(\|x - x_j\|). \quad (9)$$

Substituting Eq. (8) into Eq. (9) yields

$$u_p = \Phi A_\phi^{-1}\left[f(x) + \rho\{A_\phi\}A_\phi^{-1}u\right], \quad (10)$$

where $\Phi$ is a known matrix comprised of

$\varphi(r_{ij})$. In this study, the approximate particular solution $\varphi$ is determined beforehand, and then we evaluate the corresponding RBF $\phi$ through substituting the specified $\varphi$ into the Helmholtz equation. For the multiquadratic (MQ) RBF, the chosen approximate particular solution is

$$\varphi(r_j) = (r_j^2 + c_j^2)^{3/2}, \quad (11)$$

where $c_j$ is the shape parameter. The corresponding RBF is

$$\phi(r_j) = 6(r_j^2 + c_j^2) + \frac{3r^2}{\sqrt{r_j^2 + c_j^2}} + (r_j^2 + c_j^2)^{3/2}, \quad (12)$$

## 3. Boundary RBF Using Non-Singular General Solution

This section deals with the analogization of homogeneous solution by the RBF using non-singular general solution. We take Helmholtz operator as an illustrative example here. There exist non-singular general solutions of the other differential operators [4]. The reason for this choice is that the Helmholtz operator is the simplest among various often-encountered operators having non-singular general solution.

In the standard BEM and MFS, the Hankel function

$$H(r) = J_0(r) + i Y_0(r) \quad (13)$$

is applied as the fundamental solution of the 2D homogeneous Helmholtz equations, where $J_0(r)$ and $Y_0(r)$ are the zero-order Bessel functions of the first and second kinds, respectively. It is noted that $Y_0(r)$ has logarithm singularity, which causes major troubles applying the BEM and MFS.

The present BKM scheme discards the singular general solution $Y_0(r)$ and only use $J_0(r)$ as the radial basis function to collocate the boundary condition equations. Unlike the MFS, all nodes are placed only on physical boundary and can be used as either source or response points. The homogeneous solution of Eq. (5) is collocated by

$$v(x) = \sum_{k=1}^{N} \lambda_k J_0(r_k), \quad (14)$$

where $r_k = \|x - x_k\|$. $k$ is index of source points on boundary. $N$ is the total number of boundary knots. $\lambda_k$ are the desired coefficients. In terms of the collocation of Eqs. (6a) and (6b), we have

$$\sum_{k=1}^{N} \lambda_k J_0(r_{ik}) = D(x_i) - u_p(x_i), \quad (15)$$

$$\sum_{k=1}^{N} \lambda_k \frac{\partial J_0(r_{jk})}{\partial n} = N(x_j) - \frac{\partial u_p(x_j)}{\partial n}, \quad (16)$$

where $i$ and $j$ indicate boundary response knots respectively located on $S_u$ and $S_\Gamma$. If internal nodes are used, the following equations at interior knots are supplemented

$$\sum_{k=1}^{N} \lambda_k J_0(r_{lk}) = u_l - u_p(x_l), \quad l = 1, \ldots, L, \quad (17)$$

where $L$ is the total number of interior points used. Substituting Eq. (10) into Eqs. (15), (16) and (17), we get $N+L$ resulting simultaneous algebraic equations. It is stressed that the use of interior points is not necessary in the BKM. If only boundary knots are employed, Eq. (17) should be omitted. Chen and Tanaka [4] lists some useful non-singular general solutions. For example, it is $v(r) = \sin(r)/r$ for 3D Helmholtz problems.

## 4. Numerical Results

The geometry of two test problems is an ellipse with semi-major axis of length 2 and semi-minor axis of length 1 [5]. These examples are chosen since their analytical solutions are obtainable to compare. More complex problems can be handled in the same BKM fashion without any extra difficulty. It is stressed that only boundary knots are employed in these numerical experiments.

The inhomogeneous 2D Helmholtz problem is governed by equation

$$\nabla^2 u + u = x. \quad (18)$$

Inhomogeneous boundary condition

$$u = \sin x + x \quad (19)$$

is posed. It is obvious that Eq. (19) is also a particular solution of Eq. (18). Numerical results by the present BKM is displayed in Table 1.

Table 1. Results for Helmholtz equation

| x | y | Exact | BKM (5) | BKM (7) |
|---|---|---|---|---|
| 1.5 | 0.0 | 2.50 | 2.45 | 2.51 |
| 1.2 | -0.35 | 2.13 | 2.08 | 2.14 |
| 0.6 | -0.45 | 1.16 | 1.18 | 1.16 |
| 0.0 | 0.0 | 0.0 | 0.1 | -0.002 |
| 0.9 | 0.0 | 1.68 | 1.66 | 1.69 |
| 0.3 | 0.0 | 0.60 | 0.64 | 0.60 |
| 0.0 | 0.0 | 0.0 | 0.08 | -0.001 |

The numbers in brackets of Table 1 mean the total nodes. The shape parameter $c$ in the MQ is chosen 3. It is found that the present BKM converges very quickly. This shows the BKM holds the super-convergent merit. The BKM can yield accurate solutions with only 7 knots. In contrast, the DRBEM employs 16 boundary and 17 interior points to achieve slightly less accurate solutions for a simpler homogeneous case [5]. This is because the normal BEM has only low order of convergence speed [2].

An amazing finding of this research is that if only boundary knots are used, the BKM can formulate linear analogization equations for nonlinear problems which have nonlinear governing equations and linear boundary conditions. Consider the following case

$$\nabla^2 u + u^2 = ye^x + y^2 e^{2x}. \quad (20)$$

with inhomogeneous boundary condition

$$u = ye^x \quad (21)$$

Eq. (21) is also a particular solution of Eq. (20). If using only boundary nodes, the BKM linear formulation can be solved easily by any linear solver. Table 2 lists the BKM results compared with analytical solutions. The shape parameter $c$ in the MQ is taken 18. It is observed that the accuracy of solutions is acceptable in engineering. Compared with iteration solution of nonlinear formulation equations, the linear formulation of nonlinear problems circumvents computationally expensive task to repeatedly solve equations and the perplexing issues relating to the guess of initial solution and stability.

Table 2. Results for nonlinear equation (20)

| x | y | Exact | BKM (7) | BKM(9) |
|---|---|---|---|---|
| 4.5 | 0.0 | 0.0 | 0.0 | 0.0 |
| 4.2 | -0.35 | -23.34 | -22.78 | -23.52 |
| 3.6 | -0.45 | -16.47 | -17.10 | -16.87 |
| 3.0 | -0.45 | -9.04 | -9.60 | -9.38 |
| 2.4 | -0.45 | -4.96 | -5.12 | -5.20 |
| 1.8 | -0.35 | -2.12 | -2.01 | -2.26 |
| 3.0 | 0.5 | 10.04 | 10.66 | 10.38 |
| 3.0 | -0.5 | -10.04 | -10.66 | -10.38 |

## 5. Completeness in BKM and RBFs in Domain Collocation

Although numerical experiments show that the BKM produced accurate solutions, the possible incompleteness due to only use of the non-singular part of fundamental solution is still a concern. We only numerically tested the interior Helmholtz-type problems. For completeness, we will exam the exterior problems in later experiments. In addition, there still exist some controversies in the choice of basis functions. For example, consider homogeneous biharmonic equation

$$\nabla^4 w = 0, \quad (22)$$

we have four general solutions, namely,

$$w^*(r) = C_1 \ln(r) + C_2 r^2 \ln(r) + C_3 r^2 + C_4. \quad (23)$$

It is common practice in the BEM to use $r^2 \ln(r)$ as the fundamental solution. We are wondering if the other three general solutions will work in the framework of the MFS, BEM or the BKM. Although Duchon [6] proved that $r^2 \ln(r)$ is optimal interpolants for biharmonic operator with linear terms constraints, only use of $r^2 \ln(r)$ is insufficient in the MFS and BKM which require two independent RBFs. Some numerical and theoretical investigations should

be carried out. In the MFS, we suggest to use the following two RBFs for completeness:

$$w(r) = A(\ln(r)+1) + Br^2(\ln(r)+1), \qquad (24)$$

where *A* and *B* are undecided coefficients in the RBF.

So far the MQ is only the known RBF with exponential convergence merit. However, the troublesome shape parameter degrades its practical attractiveness. Success applying non-singular general solution in the BKM also hints that it is feasible to develop some operator-dependent RBFs for domain-type collocation methods and the DRM within the BKM, which at least partly satisfy the intrinsic characteristics of the targeted problems. It is expected that operator-dependent RBFs can also achieve superconvergence just as we found in the BKM. The following outlines novel general solution RBFs methodology.

By Green's second theorem in the indirect BEM, we have solution of Eqs. (1)-(3)

$$u(z) = \int_V G(z,x)f(x)dV(x) + \int_S \left[u\frac{\partial G(z,x)}{\partial n(x)} - G(z,x)\frac{\partial u(x)}{\partial n(x)}\right]dS(x), \qquad (25)$$

where *G* denotes the fundamental solution of operator $L\{\}-\rho\{\}$, *x* indicates source point. The above formula (25) suggests us that the RBFs can be created for interior source points by

$$\phi(r,x) = [f(x) + \rho(g(r))]r^{2m}g(r), \qquad (26a)$$

where *g(r)* is the general solution of operator. *x* represents source point coordinates. $\rho(g(r))$ in formula (26a) may be removed in some cases. *m* is integral number and $r^{2m}$ term ensures sufficient degree of differential continuity. If we place source and response nodes distinctly to avoid possible singularity, $r^{2m}$ becomes unnecessary in the presented RBFs. Formula (26a) can roughly be interpreted as transforming forcing function *f(x)* based on eigenfunctions of a operator, which has close relation with the general solution. For Dirichlet and Neumann boundary source points, we respectively have RBFs

$$\phi(r,x) = D(x)r^{2m}\frac{\partial g(r)}{\partial n} \qquad (26b)$$

and

$$\phi(r,x) = N(x)r^{2m}g(r). \qquad (26c)$$

Normal derivative in Eq. (26b) can be simply replaced by $\partial g(r)/\partial r$. For simplicity, the following RBF is also suggested

$$\phi(r,x) = r^{2m}g(r) \qquad (27)$$

for all source points. Thin plate spline (TPS) RBF can be regarded as a special case of Eq. (27) without considering inhomogeneous terms. In addition, we can construct pre-wavelet RBFs by substituting $\sqrt{r_j^2 + c_j^2}$ into Eqs. (26a,b,c) and (27) instead of *r*, where $c_j$ is dilution parameters. For example, numerical experiments using $r_j^{2m}\ln\sqrt{r_j^2+c_j^2}$ manifest spectral convergence as in the MQ. Such wavelet RBFs will be especially attractive for multiscale problems.

The RBF representation is stated as

$$u(x_i) = \sum_{k=1}^{N}\beta_k\phi(r_{ik}, x_k) + \beta_{N+1}\psi(x_i). \qquad (28)$$

$\psi(x)$ is usually the linear constraint in literature. Alternatively $\psi(x)$ represents inhomogeneous function *f(x)*, *N*(x) or *D*(x) here respectively for different response nodes located on *V*, $x \subset S_u$ or $x \subset S_T$. The side condition is

$$\sum_{k=1}^{N}\beta_k\psi(x_k) = 0, \qquad (29)$$

which assures all specified RBFs are orthogonal to inhomogeneous functions. Eq. (28) is akin to the successive approximate solution of Fredholm integral equation closely relating to the BEM, which also provides clue to prove the solvability of the present RBF interpolants system. Numerical experiments of new RBFs will be provided in a sequent paper.

Another interesting extension of the RBF concept is to introduce time-space RBFs (TSR) for time-dependent problems by considering generalized time-space field. To clearly state our idea, let us consider the equation governing wave propagation

$$u_{xx} = \frac{1}{c^2} u_{tt} + f(x,t). \qquad (30)$$

It can be restated as

$$u_{xx} + u_{tt} = \left(1 + \frac{1}{c^2}\right) u_{tt} + f(x,t). \qquad (31)$$

In this way, we can construct the RBF by means of the GSR

$$\phi(r) = r^{2m} g(r) f(x,t), \qquad (32)$$

where $g(r)$ is the general solution of extended Laplace operator of the left side of Eq. (32), the distance variable $r$ is defined as

$$r_j = \sqrt{(x - x_j)^2 + (t - t_j)^2}. \qquad (33)$$

Such definition of $r$ differs from the standard radial basis function in that the time variable is included into the distance function and is handled equally as the other space variables.

If we hope to use non-singular general solution, Eq. (30) should be rewritten as

$$u_{xx} + u_{tt} + u = \left(1 + \frac{1}{c^2}\right) u_{tt} + u + f(x,t). \qquad (34)$$

The left side of Eq. (34) is the extended Helmholtz operator including time derivative. By using the GSR, we can create the corresponding RBF

$$\phi(r) = h(r)\left[f(x,t) + h(r) + \left(1 + 1/c^2\right) h_{tt}(r)\right], \quad (35)$$

where $h(r)$ is the non-singular general solution of the extended Helmholtz operator. Another method eliminating time dependence is to use time-dependent general solutions, namely,

$$\phi(r,x,t) = t^{2m} g(r,t) f(x,t), \qquad (36)$$

where $t^{2m}$ is to avoid the possible singularity of transient general solution $g(r;t)$.

It is worth pointing out that the conventional RBF collocation method has very close relationship with the indirect BEM and the method of moments often used in electromagnetic wave propagation. Some concepts such as kernel function, source and dipole source density, and single-and double-layer potentials are useful to construct efficient RBFs. The RBF expansion coefficients in Eq. (28) can be interpreted parallel to fictitious source-density distribution in the indirect BEM.

The popular RBF schemes are mostly indirect where expansion coefficients are directly computed rather than the practical physical values. We are worried that the indirect methods may suffer troubles for problems with sharp corners. At such a corner, the expansion coefficients may tend to infinity as in the indirect BEM for interior cases, even if the practical quantities are finite. The direct RBF methods which use physical values as basic variables can avoid these problems. It is not a difficult task to convert the indirect RBF methods into direct ones by quasi-interpolation. The strategy developing direct collocation methods such as differential quadrature method can also be adopted. The direct RBF schemes also have advantage to nonlinear problems, where the practical significance of physical unknowns is helpful to guess initial iterative solution.

The globally-supported domain-type RBF methods often encounters severe ill-conditioning problems. It is noted that one source point in the RBF fashion is restricted within a certain region where it can impose significant affect on the other points. In fact, the collocation scheme can be understood as the maximum order finite difference method (FDM) of global support. Therefore, it is practically attainable to truncate the global support into finite support RBF in a way like the FDM. Consequently, we have a banded sparse system matrix of resulting equations. The difference between the FRM and the truncated MQ method [7] is that the RBF in the FRM is abruptly truncated within a certain number of neighboring nodes without using so-called decay function. Experimenting this FRM with some examples is encouraging. The

FRM also differs the compactly-supported RBF method in that any workable RBFs can be truncated to form the meshless local FRM. In addition, as in the BEM and the RBF interpolation of geometry generation problem [8], the multipole and wavelet approaches can also be used to greatly improve computing efficiency and to eliminate the ill-conditioning of the global support collocation RBF schemes.

## 6. Conclusions

The present BKM can be considered one kind of the Trefftz method [9] with features of the RBF using non-singular general solution. The method shows that singularity is not an essential ingredient in the boundary-only techniques. Comparing to the indirect BEM, this note presents general solution RBFs for various inhomogeneous problems in the domain-type RBF and DRM approximations, which fully exploit features of certain problems. Golberg and Chen [2] put the DRM on the solid RBF theory. The present work further establishes direct relationship between the RBF itself and the BEM. It is stressed that the RBF schemes may be especially attractive for higher dimension problems due to their dimensionally increased order of convergent rate [7], which makes it circumvent dimension curse.

## Acknowledgements

The authors express grateful acknowledgments of helpful discussions with Profs. C.S. Chen, M. Golberg. Y.C. Hon, and J.H. He.

## References

1. Nardini, D. and Brebbia, C.A., A new approach to free vibration analysis using boundary elements, *Applied Mathematical Modeling*, **7**(1983), 157-162.
2. Golberg, M.A. and Chen, C.S., The method of fundamental solutions for potential, Helmholtz and diffusion problems. In *Boundary Integral Methods - Numerical and Mathematical Aspects,* (M.A. Golberg, Ed), 103-176, Comput. Mech. Publ., 1998.
3. Kitagawa, T., Asymptotic stability of the fundamental solution method. *Journal of Computational and Applied Mathematics*, **38**(1991), 263-269.
4. Chen, W. and Tanaka, M., Boundary knot method: A meshless, exponential convergence, integration-free, and boundary-only RBF technique, (submitted), 2000.
5. Partridge, P.W., Brebbia, C.A. and Wrobel, L.W., *The Dual Reciprocity Boundary Element Method*, Comput. Mech. Publ., 1992.
6. Duchon, J., Splines minimizing rotation invariant semi-norms in Sobolov spaces, in: "Constructive Theory of Functions of Several Variables", Springer-Verlag, Berlin, 1976.
7. Kansa, E.J. and Hon, Y.C., Circumventing the ill-conditioning problem with multiquadric radial basis functions: applications to elliptic partial differential equations, *Comput. Math. Appls*., **39** (2000) 123-137.
8. Beatson, R.K. and Newsam, G.N., Fast evaluation of radial basis functions: Moment-based methods, SIAM J. Sci. Comput., **19** (1998), 1428-1449.
9. Piltner, R., Recent development in the Trefftz method for finite element and boundary element application, *Advances in Engineering Software,* **2** (1995), 107-115